\documentclass[rmp,twocolumn,groupedaddress]{revtex4}
\usepackage{graphics}
\usepackage{color}

\begin{document}


\title{Modeling Amyloid $\beta$ Peptide Insertion into Lipid Bilayers}

\author{David L. Mobley}
\author{Daniel L. Cox}
\author{Rajiv R.P. Singh}
\affiliation{Department of Physics, University of California, Davis, CA 95616}
\author{Michael W. Maddox}
\affiliation{Department of Chemical Engineering and Materials Science, University of California, Davis, CA 95616}
\author{Marjorie L. Longo}
\affiliation{Department of Chemical Engineering and Materials Science, University of California, Davis, CA 95616} 

\date{\today}

\begin{abstract}
Inspired by recent suggestions that the Alzheimer's amyloid $\beta$ peptide (A$\beta$) can insert into cell membranes and form harmful ion channels, we model insertion of the 40 and 42 residue forms of the peptide into cell membranes using a Monte Carlo code which is specific at the amino acid level. We examine insertion of the regular A$\beta$ peptide as well as mutants causing familial Alzheimer's disease, and find that all but one of the mutants change the insertion behavior by causing the peptide to spend more simulation steps in only one leaflet of the bilayer. We also find that A$\beta 42$, because of the extra hydrophobic residues relative to A$\beta 40$, is more likely to adopt this conformation than A$\beta 40$ in both wild-type and mutant forms. We argue qualitatively why these effects happen. Here, we present our results and develop the hypothesis that this partial insertion increases the probability of harmful channel formation. This hypothesis can partly explain why these mutations are neurotoxic simply due to peptide insertion behavior. We further apply this model to various artificial A$\beta$ mutants which have been examined experimentally, and offer testable experimental predictions contrasting the roles of aggregation and insertion with regard to toxicity of A$\beta$ mutants. These can be used through further experiments to test our hypothesis. 
\end{abstract}

\maketitle
\section{Introduction}

Scientific and public interest in Alzheimer's disease has surged in the last several decades. The reason for this is simple: with increasing life expectancy, Alzheimer's disease has emerged as the most prevalent form of late-life mental failure in humans~\cite{kn:Selkoe}.

Alzheimer's disease (AD) is a neurodegenerative disease involving progressive memory impairment, altered behavior, decline in language function, disordered cognitive function, eventual decline in motor function, and, finally, death~\cite{kn:Selkoe}. In AD, the brain is typically marked by lesions~\cite{kn:Selkoe}, neuronal damage, and vascular damage~\cite{kn:Durell}. These lesions are typically associated with extracellular plaques, called amyloid plaques, and intraneuronal fibrillar tangles~\cite{kn:Selkoe, kn:Durell}. The tangles are composed of a protein called Tau and are called Tau tangles, while the extracellular plaques are largely composed of amyloid $\beta$ peptide (A$\beta$) in 40 and 42 residue forms~\cite{kn:Selkoe} (denoted A$\beta 40$ and A$\beta 42$, respectively). These insoluble amyloid plaques composed of A$\beta$ are considered a hallmark of AD. However, they are not specific to AD~\cite{kn:Dickson} and have been observed in older patients free from AD symptoms~\cite{kn:Jarrett}. It has been pointed out that correlations between amyloid plaque density and severity of dementia are weak, while there are stronger correlations between soluble A$\beta$ levels and severity of dementia~\cite{kn:Walsh}. This is one reason for the suggestion that oligomers of A$\beta$ may be more important to toxicity than large insoluble aggregates or plaques. Evidence for this idea has been provided \emph{in vivo}~\cite{kn:Walsh} and \emph{in vitro}~\cite{kn:Lambert, kn:Hartley}. 
 
One mechanism by which oligomers can damage cells is formation of pores or ion channels through the cell membrane. Early work in this area showed that A$\beta$ can insert into model planar lipid bilayers and allow a calcium current upon insertion, and further that these channels can be blocked~\cite{kn:Arispe}, suggesting that the calcium current is really due to channel formation, not just bilayer permeabilization by the peptide. Theoretical modeling based on predicted secondary structures for membrane-bound A$\beta$ has suggested that the A$\beta$ peptide can form channels with four or six A$\beta$ subunits in each leaflet of the bilayer (for a total of 8 or 12 per channel) ~\cite{kn:Durell}. More recent work has been done using atomic force microscopy (AFM) to look at the structure of A$\beta$ inserted in planar lipid bilayers and has found what appear to be channels consisting of four or six visible subunits around a central pore, consistent with the theoretical picture described above. The monomers oligomerize after insertion into the bilayer. Furthermore, in the presence of these oligomers, current can flow~\cite{kn:Lin}. \textcite{kn:Lin} also show that, under similar conditions, A$\beta 42$ induces neuritic degeneration and death in cell culture and that this toxicity is calcium-dependent and blocked by zinc. Imaging work by another group has also shown that $A\beta 40$ oligomers with the E22G mutation (where glutamate (E) at residue 22 is replaced with glycine (G)), which causes a form of familial AD, can form pore-like structures~\cite{kn:Lashuel}. These pore-like structures actually could be intermediates which, when not membrane bound, build up into the amyloid plaques observed in the brain of AD patients~\cite{kn:Lashuel}. 

Based on these suggestions, and the observation of \textcite{kn:Lin} that oligomers in the membrane form after insertion of monomers, we model insertion of the A$\beta$ peptide into the cell membrane. We first examine the regular A$\beta 40$ and A$\beta 42$ peptides, then the 40 and 42 residue versions of all of the mutations in the A$\beta$ peptide that are known to cause familial AD (FAD) and reduce the average age of onset for the disease compared to people with sporadic AD~\cite{kn:Selkoe}. We believe FAD mutants provide a tool for assessing proposed toxicity mechanisms, in that the biological toxicity mechanism should explain why these mutants cause FAD. Our reasoning in looking at these mutants is that if the insertion behavior of the FAD mutant peptides is different, this could make a difference in the prevalence of oligomers in the membrane and thus have an effect on toxicity, if membrane-associated oligomers are indeed important for toxicity \emph{in vivo}. While some earlier modeling work has dealt with the structure of A$\beta 40$ in a lipid bilayer~\cite{kn:Pellegrini}, we believe this work is the first to compare insertion of FAD mutants.

This system is modeled using a Monte Carlo (MC) code which has been developed to study insertion behavior of peptides into lipid bilayers. This model, which is specific at the amino acid level, allows us to simulate larger peptides and longer timescales than traditional molecular dynamics simulation studies. The configurational steps are sufficiently small that it has been used successfully to suggest insertion mechanisms, as well as to describe insertion conformations for some peptides~\cite{kn:Maddox, kn:Maddox2}. Here, we find that in all cases the peptide inserts relatively easily. However, we find differences in the conformations the peptide adopts once inserted. These differences in the prevalence of conformations are our central result. Relative to the normal A$\beta$ peptide, most of the FAD mutant peptides are more likely to insert only partially in the bilayer. We point out similarities between this partially inserted conformation and the predicted channel structures~\cite{kn:Durell}. Thus we suggest that FAD mutants may, in this way, facilitate formation of harmful channels. Moreover, the A$\beta 42$ peptide, with additional hydrophobic residues, has a greater tendency than A$\beta 40$ to hang up in this conformation, and this may correlate with the increased toxicity of A$\beta 42$.

\section{Introduction to the model and method}

\subsection{Model energy function}
The Monte Carlo model used here has been described in detail in an earlier publication in this journal~\cite{kn:Maddox}. Accordingly, we give a brief overview of the essentials here and direct the interested reader to the earlier reference for greater detail.
 
The model follows previous work from the past decade, most notably and closely that of Milik and Skolnick~\cite{kn:MilikSkolnick1, kn:MilikSkolnick2} and Baumgaertner\cite{kn:Baumgaertner}.  Each amino acid residue is treated as a sphere of identical 1.5\AA~ radius.  There are three contributions to the potential energy function which are residue independent: 1) $U_s$, which is a hard core steric interaction preventing residue-residue overlap; 2) $U_T$, an energy measuring the cost of rotating the peptide planes of successive residues, which is periodic in the torsional angle $\phi$ between successive residues and has a shallow minimum at $\phi=52.1^o$; and 3) $U_A$, characterizing the energy of distortion of the angle $\theta$ between adjacent bonds, with a shallow minimum at $\theta=89.5^o$.

The lipid bilayer with surrounding water is modeled as a medium without molecular specificity, but with three different spatial regions. The bilayer's normal is taken to lie along the $z$-axis, so these regions are invariant in the $x-y$ plane. The bilayer has overall thickness $2(z_0 + z_h)$, where $z_0$ is the length of the acyl chains of a given leaflet, and $z_h$ is the width of the headgroup region.  We have used $z_h=$4.5\AA~ and $z_0=$13.5\AA. We have also tried different chain lengths $z_0$ but we do not present the results here as they were not significantly different except for reductions in the amount of the transbilayer conformation of the inserted peptide when the membrane is sufficiently thick, as we discuss in the results section. 

The water-lipid medium is characterized by three dimensionless functions, two of which couple linearly to residue specific parameters we discuss in the next paragraph.  These functions are (as shown in Fig. 1):

\begin{figure}
\includegraphics{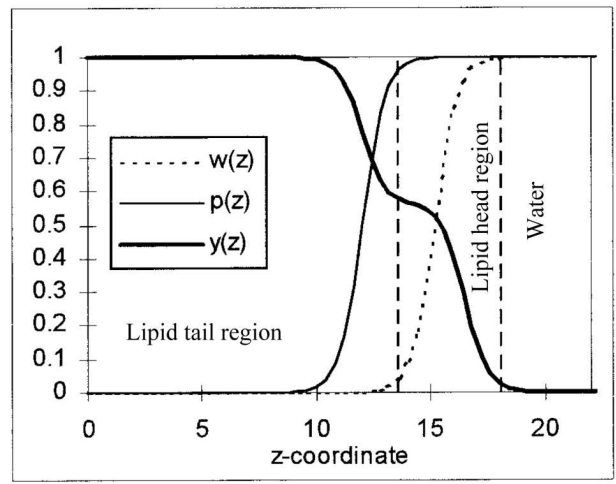}
\caption{{\bf Functions characterizing bilayer properties.} The lipid bilayer is described by three functions, $w(z)$ for the fractional water content, $p(z)$ for the polarity, and $y(z)$ for the hydrophobicity. Here the z-axis is perpendicular to the plane of the bilayer, and the functions (and the bilayer) are symmetric around z=0.}
\end{figure}

1) $w(z)$, which measures the fractional water content; this is modeled as a step function with exponentially rounded edges (decay length of 2\AA) that is zero in the hydrophobic acyl chains, one in the water region, and varies smoothly through the head region.

{2) A polarity function $p(z)$, also exponentially rounded with the same decay length, and chosen to be one in the lipid head regions and water while falling to some small value $1-f_q$ (where $f_q$, the polarity factor, determines the polarity of the tail region, with larger $f_q$ corresponding to a less polar tail region) after approximately one residue diameter into the tail region. 

3) A hydrophobicity function $y(z)$, which is the sum of two exponentially rounded step functions: one which is zero in the water region and saturates in the head region, proportional to the total gain of hydrophobic energy in the head region, and a second which saturates in the tail region after approximately one residue diameter and accounts for the hydrophobic energy gained for residues penetrating the acyl tail region. 

The water content $w(z)$ couples linearly to the external hydrogen bonding energy of each residue, which is residue independent in form.  The net hydrogen bonding energy is taken as $U_H$ given by a sum over residues $i$

\begin{equation}
U_H = \sum_i(w(z_i)H_0 +(1-w(z_i))H_{int}(i))
\end{equation}
where $H_0=-6.12$ kcal/mol is the transfer energy of an unbonded peptide group to water and $H_{int}(i)$ the internal hydrogen bonding energy associated with $\alpha$-helix formation given by 
\begin{equation}
H_{int}(i) = {H_0\over 4} \sum_{n=-4,-3,3,4}V_H(|\vec r_{n+i}-\vec r_i|)~~.
\end{equation}
$V_H$ is a nearly hard-core function of the separation between residue $i$ and the potential helical hydrogen bonding partners along the peptide chain, as proposed by Milik and Skolnick~\cite{kn:MilikSkolnick2}.

Residue specificity is included in two energies associated with polarity and hydrophobicity.  First, a potential energy term $U_Q = \sum_{i} q_0(i)p(z_i) $ is included, where $q_0(i)$ is the residue-specific polar energy associated with charged or partially charged functional groups. Second, a hydrophobic energy $U_B = \sum_i B(i)$ is included, where
\begin{equation}
B(i) = y(z_i)b_0(i) + (1-y(z_i)){b_1(i)\over 4}\sum_{n=-4,-3,3,4}V_H(|\vec r_{n+i}-\vec r_i|)~~.
\end{equation}
Here, $b_0(i)$ is residue specific and measures the water-to-alkane Gibbs hydrophobic transfer energy for residue $i$, and $b_1(i)$ is the maximum reduction in hydrophobic energy due to helical folding.  $b_0(i)$ is taken to be proportional to the stochastic accessible area of the residue.  The helical-folding related term derives from the loss of accessible surface area associated with helix formation.  Values for $q_0(i),~ b_0(i)$, and 
$b_1(i)$ for all residues are tabulated in the previous work~\cite{kn:Maddox}. Note that because the model treats the lipids and water only as media, the hydrophobic energy must be included explicitly in our model energy function.
 
Because of the way the hydrophobic and hydrogen bonding energies are calculated -- simply based on local helicity -- the simulation is biased toward alpha helices, as beta structure involves longer range interactions and is not taken into account by the model. Therefore, the model will not accurately describe insertion behavior of any peptide that inserts while in a conformation rich in beta structure. Fortunately, the monomeric A$\beta$ peptide is predicted, based on secondary structure, to be alpha helical between residues 15 and 40 or 42~\cite{kn:Durell,kn:Pellegrini} when membrane-bound. Experimental NMR work in aqueous sodium dodecyl sulfate micelles, which to some extent resemble a water-membrane medium, confirms this for A$\beta 40$~\cite{kn:Coles}. Thus the model's bias away from beta structure should not play a significant role here. Indeed, we find that the region mentioned above inserts into the membrane in a largely helical structure, as described below.

Our total energy is then taken as the sum $U=U_S+U_T+U_A+U_B+U_Q+U_H$.  
All of our modeling presented below is done at pH 7.0 with temperature 305 K, and uses a polarity factor $f_q=0.85$, corresponding to a polarity between that of octanol and hexadecane. The choice of this value is based on experimental studies~\cite{kn:Roseman, kn:Griffith} and earlier simulation work~\cite{kn:Maddox}.

\subsection{Monte Carlo simulation details} 

The simulation method is the canonical MC method. We use periodic boundary conditions in all three directions, and in the case where the peptide runs across the boundary, interactions are calculated using the minimum separation between the two residues in question (the minimum image convention). New peptide conformations are generated using three different sorts of moves:

1) Peptide translation: The whole peptide is randomly translated a small distance (between 0 and 0.2 \AA ) along each Cartesian axis.

2) Spike move (two sorts): (a) For an end residue, the virtual bond connecting it to the chain is rotated slightly, first in the x-y plane, and then in the y-z plane. The angle of rotation is random, between $0^o$ and $20^o$. (b) For a central residue, the residue is rotated a random (between $0^o$ and $20^o$) amount around an axis joining the centers of its nearest neighbors, while keeping all virtual bond lengths fixed.

3) Slide move: A random virtual bond is selected and all residues on one side (selected randomly, either up or down the chain) of it are moved a small, random amount (between 0 and 0.2 \AA ), while remaining fixed relative to one another. The move leaves the initial virtual bond the same length but rotated relative to the residues it connects.

One MC step consists of one modification of each type 1, 2, and 3, where the choice of residue is random for (2) and (3). Modifications are accepted or rejected with a probability given by the usual Boltzmann factor $p=e^{-\frac{\Delta U}{RT}}$ so that favorable moves, with a negative $\Delta U$, are always accepted, and some unfavorable moves are accepted.

In our work, we wanted to capture insertion behavior without biasing results by initial peptide conformations. We have done two groups of simulations to accomplish this. First, we have started the peptide outside the bilayer in the aqueous phase in a random conformation. Second, we have started the peptide in an initially helical, fully inserted conformation.

MC simulations can be used to investigate non-equilibrium properties (e.g. insertion mechanisms) or equilibrium properties (e.g. inserted conformations) of a system. When these simulations are used to study equilibrium properties, it is important to ensure that the system has fully equilibrated before data collection begins. If this is not done carefully, one consequence is that the so-called equilibrium state might depend on the initial conditions. To establish an appropriate period of equibration, we monitored the average energy of the peptide as a function of simulation steps. As the peptide equilibrates (reaching its energetically preferred conformation(s)), the average energy decreases from an initially higher value. Thus we can get a reasonable idea how many simulation steps it takes for this to happen simply by plotting energy versus step number.

Using this method, we find that for insertion from an initial conformation outside the bilayer, a 30 million step equilibration period is usually sufficient, while for an initially inserted and helical conformation, 30 million steps is always sufficient. Although both initial conformations eventually produce the same equilibrated state, the inserted helical conformation converges more rapidly and is used, with an equilibration period of 50 million steps, in all our simulations (unless otherwise noted).

As a further test that our equilibration period is sufficient, we have also used conformations from peptides at the end of an entire simulation run as starting points for new simulations, and the results at the end of both simulations are within our error bars of one another.

It is worth pointing out that no equilibration would be required if the insertion mechanism is being studied. However, in this work, we find that in every case insertion is fairly easy, as we discuss in the results section. Thus we focus on peptide conformations at equilibrium.

\subsection{Data collection}
Every MC simulation run employs a unique set of random numbers, resulting in a slightly different result each trial (similar to the way in which no two experimental measurements are identical). More accurate data are therefore generated by averaging multiple runs. 
Here, we have run a minimum of 10 trials for every peptide: five beginning in initially helical and inserted conformations, and five with initially random conformations outside the bilayer. Following equilibration (discussed above), data is collected for 50 million steps. In this paper we report the results of the initially inserted conformations, thus our results are averaged over a minimum of five such trials. However, as we will discuss in the appendix on data analysis, in some cases we use more trials.

\section{Results}
\subsection{Overview}
Overall, we simply input the sequence of A$\beta 40$ and $A\beta 42$ and various mutants and run the simulations. As discussed above, we do multiple trials for each peptide and average the results. What we find, briefly, is that the peptides all insert relatively easily into the bilayer if they begin initially outside the bilayer, and we do not find that the FAD mutations significantly effect this ease of insertion. However, we find that the mutations do influence the conformation the peptide adopts once inserted into the bilayer. Therefore, our focus in this work is not on details of how the peptide inserts, though we believe it is likely that insertion proceeds via one of the two main insertion mechanisms described previously~\cite{kn:Maddox, kn:Baumgaertner}.

One of our fundamental results is that the peptide appears to exhibit multiple possible inserted conformations which have nearly the same energies, thus allowing the peptide to switch between conformations often in the course of a simulation. We have previously described such behavior as conformational partitioning~\cite{kn:Maddox2}. We find that that the A$\beta$ peptide and its mutants can always adopt the same, small set of conformations. However, the mutations alter the number of MC steps the peptide spends in each of these conformations (which, in a real system, would correspond to the number of inserted peptides in each conformation). These conformations are shown in Fig. 2 and Fig. 3.

\begin{figure}
\includegraphics{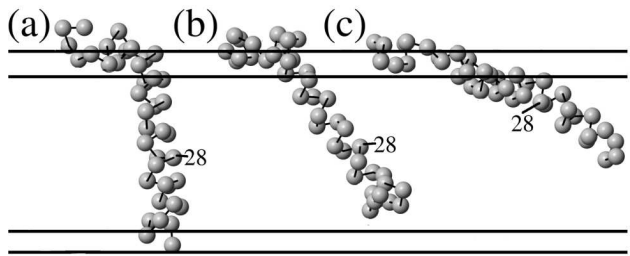}
\caption{{\bf Primary inserted conformations of the A$\beta$ peptide.} We find that in every case, the inserted peptides can adopt essentially three different conformations. Mutations appear to alter the percentage of steps the peptide spends in each conformation but do not fundamentally change the conformations. (a) Transbilayer: The peptide inserts with the last several residues near the C-terminus in the lower lipid head region; the portion crossing the bilayer is roughly helical. (b) Fully Inserted: Just like (a), except the last several residues are not anchored in the lower head region, meaning that the conformation is fairly flexible. (c) Partially Inserted: Like (b), except now much more of the peptide is tethered to the upper head region by the polar residues 22-23 and 26-28, whereas before only residues 1-15 or so were in the upper head region. The conformations shown are for A$\beta 40$, but A$\beta 42$ has similar conformations with two additional residues (isoleucine and alanine) the the C-terminus.}
\end{figure}

\begin{figure}
\includegraphics{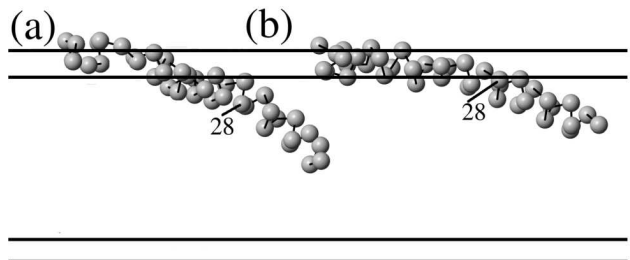}
\caption{{\bf Sub-configurations associated with locations of the polar residues.}We can further break up the partially inserted conformation, (c) from Fig. 2, into two conformations: (a) a conformation where only residues 22-23 remain in the upper head region, and (b) a conformation where both residues 22-23 and 26-28 remain in the upper head region.}
\end{figure}

Essentially, it is quite easy to distinguish three conformations: the first has the last several residues anchored in the lower head region. We call this conformation ``transbilayer''. The second conformation is similar, but usually much more prevalent since the last residues are hydrophobic and prefer to remain in the tail region. This is especially true in the case of A$\beta 42$. In this conformation, the essential difference is that the tail of the peptide is not anchored and thus is fairly floppy and able to change the angle it makes with the z-axis easily. We call this conformation ``fully inserted''. The third conformation is different in that in the first two, only residues 1-15 or so remain in the upper lipid head region, while in the third conformation, the polar residues 22-23 and/or 26-28 also remain in the upper head region (along with some of their neighbors), where there is still some water content. As a result, the C terminus (residue 40 or 42) does not stick down into the lipid tail region nearly as far as in the other two conformations. We call this conformation ``partially inserted''. In this case all, or almost all, of the peptide is only in the upper leaflet of the bilayer. 

This third, partially inserted conformation can be divided into two conformations simply by distinguishing whether it is residues 22-23 that remain in the upper head region, or also residues 26-28. This separation of conformations is shown in Fig. 3. Some of our analysis is done grouping these together, and some by separating them, as we will discuss below. 

Briefly, we find that when the FAD mutations have an effect on the insertion behavior, it is usually by causing the mutant peptides to favor the partially inserted conformation more than wild-type. The E22G mutant is an exception, as it essentially eliminates this conformation.

We refer the reader to the appendix for detailed discussion of our data analysis procedure. Overall, however, the basic output of the simulation is the number of steps, or percentage of steps, each residue in the peptide spends at each z-coordinate. We can plot this for all residues (Fig. 4) or particular residues (Figs. 5-7). With some analysis of these (the data analysis is explained in the appendix) we are able to get accurate measurements in increases or decreases in the number of steps the peptide spends in a given conformation, relative to wild-type. We are able to do this whether we choose to separate the peptide's conformations into three or four groups. Results obtained using these methods are presented in Table 1 and Table 2. 

\begin{figure}
\includegraphics{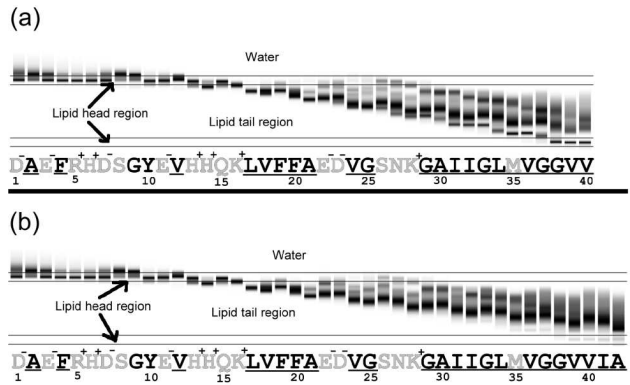}
\caption{{\bf Number of steps at each z-coordinate (vertical axis is z-axis; darker means more steps) plotted versus residue number} (sequence shown), from 1-40 for A$\beta 40$, (a), and from 1-42 for A$\beta 42$, (b). Note that on residue 30, there appear three dark regions, corresponding to three peaks, while residues 26 and 28, for example, have four peaks. This result is important for our data analysis. Note also that the transbilayer conformation for A$\beta 42$ is less common than for A$\beta 40$ (compare the darkness, or number of steps, of the lowest peak on residue 40 for A$\beta 40$ and A$\beta 42$. The sequence of A$\beta$ has polar residues in gray, hydrophobic residues underlined, and charged residues with charges indicated.}
\end{figure}

\begin{table*}
\noindent \begin{tabular}{||l|r|l|l|r|r|r|r||}
\hline
& &  & \multicolumn{2}{|c|}{Three peak analysis} & \multicolumn{2}{|c|}{Four peak analysis} \\ \cline{3-5} \cline{6-7}
Peptide form & Onset & \% trans. & \% fully ins. & \% partially ins. & $\Delta$\% upper(a) & $\Delta$\% upper(b) \\ \hline
WT A$\beta 40$	& 72.8	& 15.6 +/- 3.8 & 48 +/- 5.2 & 36.3 +/- 6.8 & & \\ \cline{2-7}
A$\beta 42$ & & 2.1 +/- 0.5	& 58.5 +/- 5.6 & 39.4 +/- 5.8 & & \\ \hline
A21G A$\beta 40$ & 52 & 17.6 +/- 3.0 & 48.3 +/- 6.8 & 34.1 +/- 8.0 & 0.5 +/- 1.8 &  2.9 +/- 8.4 \\ \cline{2-7}
A$\beta 42$ & & 1.9 +/- 0.7 & 44.4 +/- 5.4 & 53.7 +/- 5.4 & -2.8 +/- 3.2 & 14.8 +/- 6.9 \\ \hline
E22G A$\beta 40$ & 57  & 27.4 +/- 7.5 & 67.4 +/- 7.0 & 5.13 +/- 7.0 & -11.8 +/- 1.6 &  -13.2 +/- 6.2 \\ \cline{2-7}
A$\beta 42$ & & 2.3 +/- 1.7 & 94.8 +/- 7.0 & 1.29 +/- 1.0 & -16.8 +/- 3.7 & -9.6 +/- 5.6 \\ \hline
E22Q A$\beta 40$ & ?  & 13.6 +/- 4.6 & 43.6 +/- 5.5 & 42.8 +/- 5.5 & 6.4 +/- 3.5 &  2.8 +/- 11.0 \\ \cline{2-7}
A$\beta 42$ & & 2.0 +/- 0.3 & 55.8 +/- 6.9 & 42.2 +/- 7.0 & 4.5 +/- 3.4 & -1.8 +/- 8.7 \\ \hline
E22K A$\beta 40$ & ?  & 15.5 +/- 9.6 & 37.9 +/- 5.2 & 46.6 +/- 6.0 & 9.9 +/- 2.3 &  2.2 +/- 10.0 \\ \cline{2-7}
A$\beta 42$ & & 1.3 +/- 0.8 & 34.2 +/- 10.5 & 64.5 +/- 9.4 & 9.4 +/- 8.0 & 14.6 +/- 17.1 \\ \hline
D22N A$\beta 40$ & 69  & 13.1 +/- 9.2 & 30.0 +/- 8.8 & 56.8 +/- 10.2 & 8.6 +/- 7.2 &  10.8 +/- 16.2 \\ \cline{2-7}
A$\beta 42$ & & 1.2 +/- 0.5 & 41.1 +/- 16.3 & 57.7 +/- 16.7 & 0.6 +/- 5.8 & 16.0 +/- 11.4 \\ \hline

\end{tabular}

\caption{{\bf Frequency of each conformation for native and FAD A$\beta$ peptides. }Ages of onset (where known)~\cite{kn:Li, kn:Roks, kn:Nilsberth, kn:Grabowski} and simulation results:  percentage transbilayer, percentages in the fully inserted (Fig. 2(b)) and partially inserted (Fig. 2(c)) conformations as calculated from the three peak analysis on residue 30. The last two columns measure the change in percentage of the conformations of Fig. 3(a) and 3(b) relative to wild-type (from the four peak analysis). For each mutation, one row represents A$\beta 40$ and the next A$\beta 42$. Onset age is not specific to the 40- or 42- residue forms. Here, the only consistent trend that we find is that most of the FAD mutants, with the exception of E22G, appear to increase the percentage of the partially inserted conformation relative to wild-type. This appears to be true for both A$\beta 40$ and A$\beta 42$. By way of comparison, the FAD mutations increase fibrillar aggregation of A$\beta 42$ \emph{in vitro} in every case except the A21G mutant~\cite{kn:Murakami} and increase soluble A$\beta$ levels in every case but the E22G mutant~\cite{kn:Nilsberth, kn:Lashuel}. To improve statistics, wild-type A$\beta 40$ results are the average of 25 trials rather than the usual 5; wt A$\beta 42$ are the average of 20; A21G A$\beta 40$ 20 trials; A21G A$\beta 42$ 10 trials, and A$\beta 42$ E22Q and D23N 10 trials each. All the rest are 5 trials, as described in the data analysis section.}

\end{table*}

\begin{table*}
\noindent \begin{tabular}{||l|r|r|r|r|r||}
\hline
&  & \multicolumn{2}{|c|}{Three peak analysis} & \multicolumn{2}{|c|}{Four peak analysis} \\ \cline{2-4} \cline{5-6}
Peptide form & \% trans. & \% fully ins. & \% partially ins. & $\Delta$\% upper(a) & $\Delta$\% upper(b) \\ \hline
WT A$\beta 40$	& 15.6 +/- 3.8 & 48 +/- 5.2 & 36.3 +/- 6.8 & & \\ \cline{2-6}
A$\beta 42$  & 2.1 +/- 0.5	& 58.5 +/- 5.6 & 39.4 +/- 5.8 & & \\ \hline
E22QD23N A$\beta 40$ & 15.5 +/- 5.0 & 30.0 +/- 6.4 & 54.6 +/- 6.9 & 4.4 +/- 6.3 &  13.9 +/- 16.2 \\ \cline{2-6}
A$\beta 42$ & 0.9 +/- 0.7 & 24.6 +/- 11.4 & 74.4 +/- 11.4 & 4.0 +/- 10.5 & 30.9 +/- 19.0 \\ \hline
E22A A$\beta 40$ & 23.1 +/- 7.6 & 71.7 +/- 7.1 & 5.2 +/- 0.7 & -10.8 +/- 2.3 & -13.3 +/- 6.3 \\ \cline{2-6}
A$\beta 42$ & 2.0 +/- 0.4 & 85.6 +/- 2.1 & 12.4 +/- 2.0 & -13.8 +/- 7.0 & -11.4 +/- 4.2 \\ \hline
E22D A$\beta 40$ & 22.9 +/- 8.9 & 64.9 +/- 7.6 & 12.2 +/- 2.0 & -3.9 +/- 2.6 & -13.1 +/- 6.2 \\ \cline{2-6}
A$\beta 42$ & 1.9 +/- 1.1 & 62.9 +/- 6.6 & 35.2 +/- 6.6 & -5.9 +/- 3.4 & -0.4 +/- 10.7 \\ \hline
A2S A$\beta 40$ & 17.6 +/- 9.1 & 49.3 +/- 6.4 & 33.1 +/- 5.0 & -1.2 +/- 9.9 & --3.2 +/- 9.7 \\ \cline{2-6}
A$\beta 42$ & 2.2 +/- 1.3 & 60.4 +/- 3.4 & 37.4 +/- 3.3 & 2.4 +/- 3.3 & -5.3 +/- 8.7 \\ \hline
F19S A$\beta 40$ & 1.6 +/- 2.4 & 5.3 +/- 4.4 & 93.1 +/- 4.9 & 4.9 +/- 10.3 & 57.9 +/- 16.1 \\ \cline{2-6}
A$\beta 42$ & 0.1 +/- 0.1 & 10.8 +/- 5.2 & 89.1 +/- 5.2 & 7.8 +/- 13.6 & 43.6 +/- 14.9 \\ \hline
I32S A$\beta 40$ & 14.2 +/- 7.0 & 34.7 +/- 8.7 & 51.1 +/- 10.7 & -1.7 +/- 3.2 & 17.0 +/- 12.6 \\ \cline{2-6}
A$\beta 42$ & 1.5 +/- 0.7 & 49.5 +/- 13.9 & 49.1 +/- 14.1 & -1.1 +/- 4.5 & 10.2 +/- 11.5 \\ \hline
I32V A$\beta 40$ & 16.0 +/- 3.9 & 58.8 +/- 2.8 & 25.3 +/- 1.3 & 4.4 +/- 1.6 & -10.8 +/- 6.8 \\ \cline{2-6}
A$\beta 42$ & 2.5 +/- 1.2 & 63.2 +/- 2.1 & 34.4 +/- 2.0 & -0.1 +/- 3.7 & -5.7 +/- 5.6 \\ \hline
V36E A$\beta 40$ & 83.0 +/- 13.9 & 10.6 +/- 9.3 & 6.4 +/- 6.2 & -12.7 +/- 3.0 & -5.3 +/- 13.3 \\ \cline{2-6}
A$\beta 42$ & 16.7 +/- 11.9 & 36.2 +/- 19.5 & 47.1 +/- 10.5 & 1.6 +/- 5.1 & 5.6 +/- 11.8 \\ \hline
H6R A$\beta 40$ & 17.3 +/- 4.8 & 47.7 +/- 7.7 & 35.0 +/- 8.4 & 2.2 +/- 2.5 & -1.2 +/- 8.8 \\ \cline{2-6}
A$\beta 42$ & 2.0 +/- 0.6 & 52.9 +/- 8.2 & 45.1 +/- 8.4 & 1.8 +/- 3.7 & 2.5 +/- 7.3 \\ \hline

\end{tabular}

\caption{{\bf Frequency of each conformation for artificial A$\beta$ mutants. }Shown are results for artificial A$\beta$ mutants mentioned in the literature: percentage transbilayer, percentages in the fully inserted (Fig. 2(b)) and partially inserted (Fig. 2(c)) conformations as calculated from the three peak analysis on residue 30. The last two columns measure the change in percentage of the conformations of Fig. 3(a) and 3(b) relative to wild-type (from the four peak analysis). The A$\beta 40$ E22QD23N mutant is more toxic to HCSM cells than either mutant alone~\cite{kn:VanNostrand}, while the E22D mutant of A$\beta 40$ is not toxic to HCSM cells and the E22A mutant is~\cite{kn:Melchor}. The A2S, F19S, I32V, I32S, and V36E mutants are known to reduce aggregation of A$\beta 42$~\cite{kn:Wurth}, and the H6R mutant has been suggested as an FAD mutant~\cite{kn:Janssen}.}

\end{table*}

\subsection{Normal A$\beta$ peptide insertion}
As detailed in the appendix, some results for A$\beta 40$, A$\beta 42$, and various mutants, are shown in Fig. 5, 6, and 7. We have also extracted the number of steps in each conformation and present these (measured as percentages of total steps) in Table 1, along with standard deviations. As we discuss in the appendix, the percentages for conformations (b) and (c) are not completely accurate as absolute measures of the number of steps in those conformations, but the change relative to wild-type is accurate.

\begin{figure}
\includegraphics{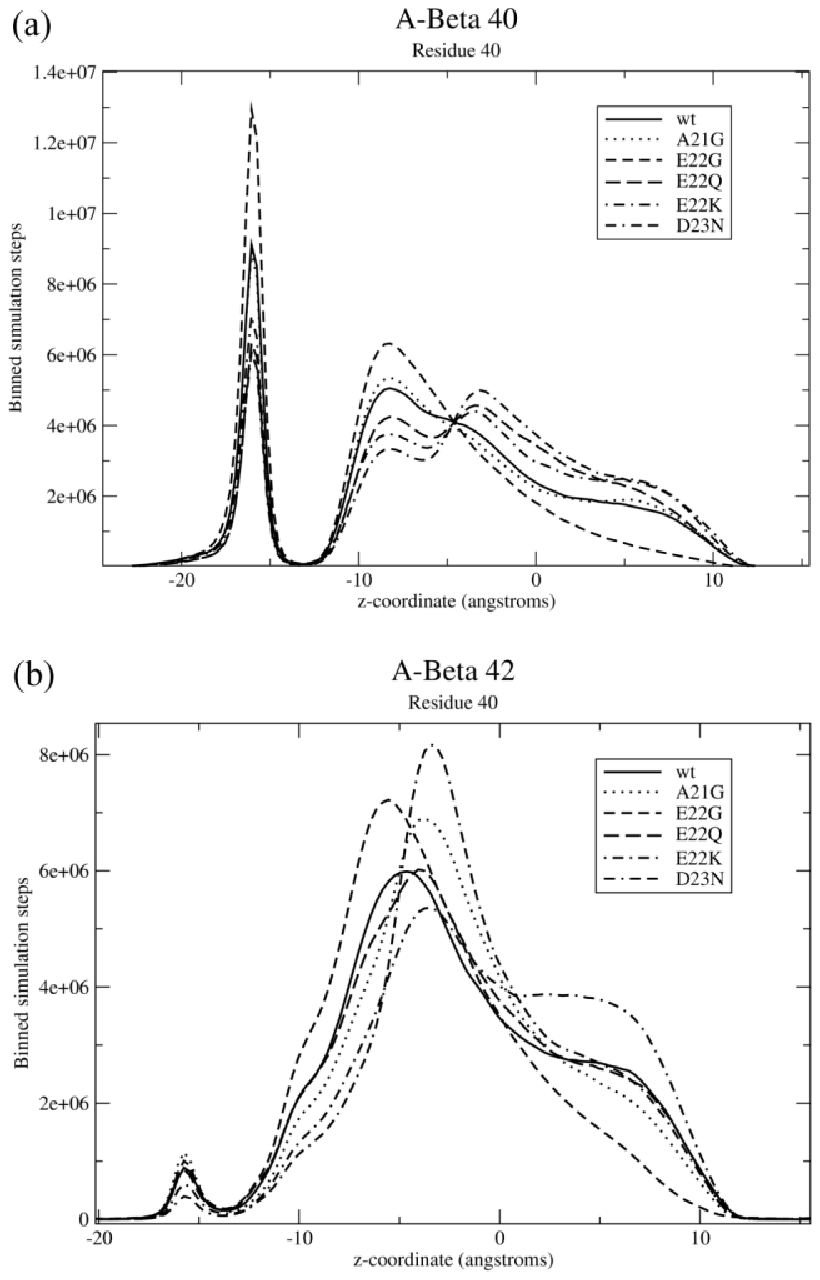}
\caption{{\bf Binned numbers of steps spent at each z-coordinate for residue 40 of (a) A$\beta 40$ and (b) A$\beta 42$}. Each plot also shows various FAD mutations. It is easily apparent that A$\beta 42$ spends significantly fewer steps in the fully inserted conformation (leftmost peak) compared to A$\beta 40$. It is difficult to tell much about the other conformations by looking at this distribution for residue 40, but the prevalence of these can be extracted from other residues. Note that the lipid head regions are from z=13.5\AA to z=18\AA (and similarly for negative z).}
\end{figure}

\begin{figure}
\includegraphics{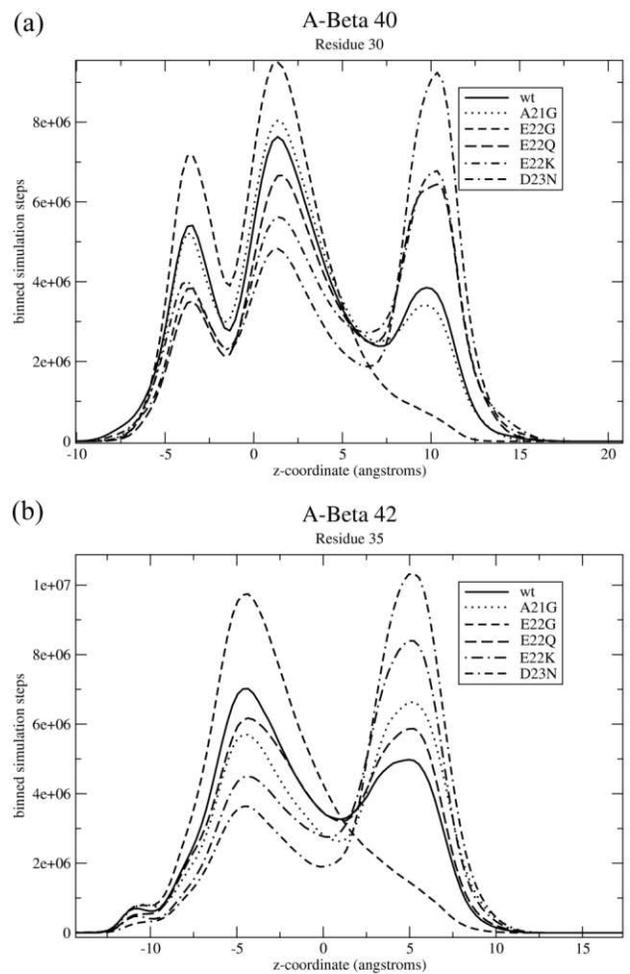}
\caption{{\bf Binned number of steps spent at each z-coordinate for (a) residue 30 of A$\beta 40$ and (b) residue 35 of A$\beta 42$}. These are the residues we picked which best distinguish three groups of conformations. The leftmost group on both, which is very small for A$\beta 42$, is conformations which appear nearly transbilayer. The middle group is conformations which are inserted and fairly floppy, as in Fig. 2(b), and the rightmost group is the partially inserted conformation. It can be clearly seen that for A$\beta 40$, all of the mutations but E22G and A21G result in an increase in this last peak relative to wild-type, and for A$\beta 42$ all of them except E22G do, as well.}
\end{figure}

\begin{figure}
\includegraphics{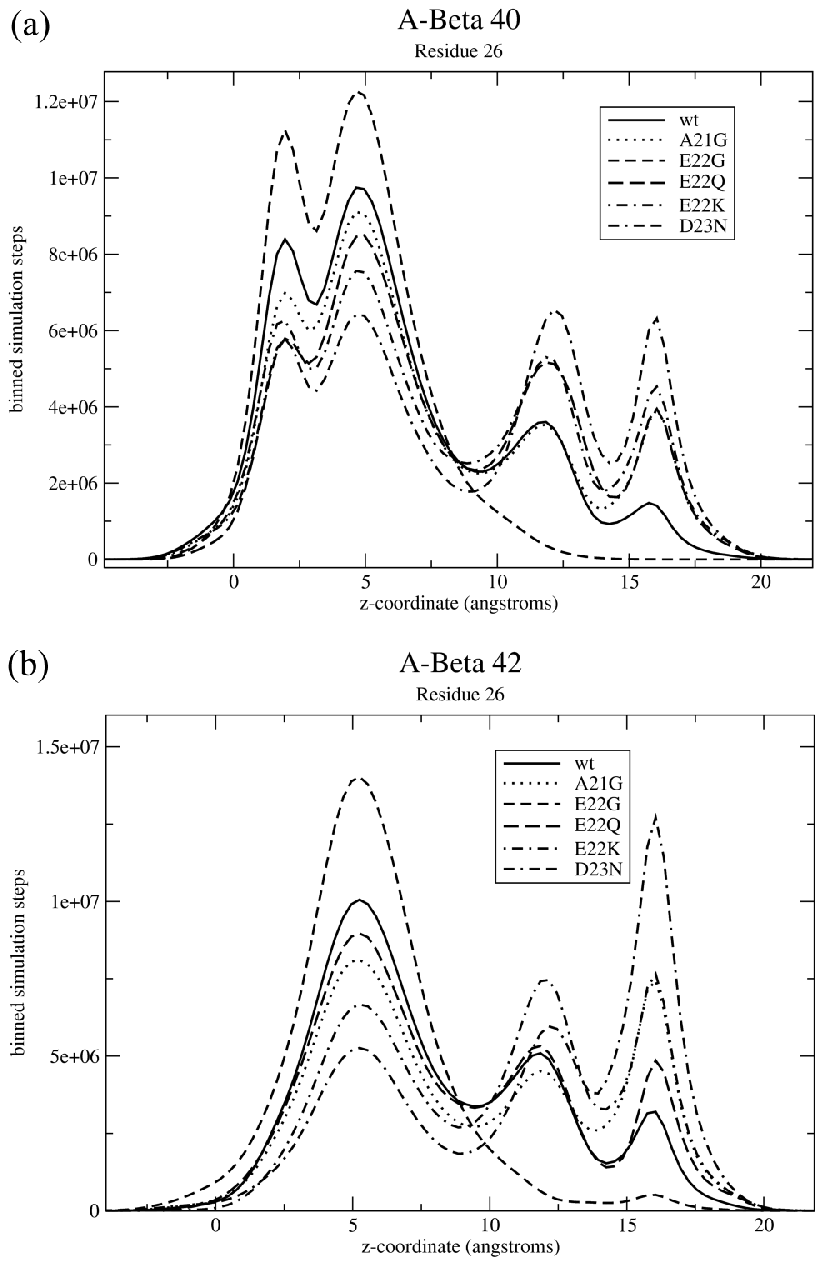}
\caption{{\bf Binned number of steps spent at each z-coordinate for residue 26 of (a) A$\beta 40$ and (b) A$\beta 42$}. For this residue, there are four apparent peaks, corresponding to the conformations of Fig. 2(a) and (b), and Fig. 3. For A$\beta 42$, the transbilayer conformation is so small that there is no apparent peak. For A$\beta 40$, it is the leftmost peak, followed by the fully inserted peak, then the partially inserted peaks: the peak of Fig. 3(a), and then the peak of 3(b). Notice that for both A$\beta 40$ and A$\beta 42$, the FAD mutants increase the weight of the rightmost conformations (those of Fig. 3) relative to wild-type, except for the E22G mutant.}
\end{figure}

It is important to note that there is a fundamental difference between A$\beta 40$ and A$\beta 42$. While both can adopt all the conformations of Fig. 2, A$\beta 42$ spends many fewer steps in conformation (a). This is because the last two residues of A$\beta 42$ add significantly to the hydrophobicity of the C-terminus (see sequence in Fig. 4), as both are nonpolar and isoleucine is strongly hydrophobic. One might well ask, however, why A$\beta 40$ has its last three residues in the lower head region at all, especially the final two valines, which are nonpolar and strongly hydrophobic. To understand this effect, it is worth noting that in this conformation, the transbilayer helix begins after, or around, residue 16, lysine, which is the final charged and polar residue prior to a number of nonpolar, hydrophobic residues, beginning with leucine. To remove residues 39 and 40 from the lower head region would require either the peptide to insert at a more shallow angle so that these residues do not make contact with the lower head region, or putting a kink in the helix (as in Fig. 2(b)) so that they do not make contact. In the first case, changing the insertion angle would move lysine, and possibly residues after it (it is immediately followed by valine), into the upper head region where there is still some water. This would be costly energetically. So is putting a kink into the helix. Thus we believe that, in the case of A$\beta 40$, the energy cost of either putting a kink in the helix, or imbedding K16 and V17 in the upper head region, is comparable to the energy cost of imbedding V39 and V40 in the lower head region, thus the conformation 2(a) does occasionally happen. On the other hand, for A$\beta 42$, the additional I41 makes this conformation so costly that it almost never happens. This reduction in the transmembrane conformation results in an increase in the prevalence of the other two conformations relative to A$\beta 40$ which, if correct, could help explain why A$\beta 42$ is typically more toxic.

It is probably important to point out that our results do not necessarily mean A$\beta 40$ would be transmembrane biologically. We have tried making the bilayer slightly thicker (by 5\AA) and the transbilayer conformation essentially disappears.  This means the transmembrane insertion is due to the limited space between head regions, as we argued above. Biologically, the membrane could be slightly thicker than our model, or the peptide might cause a small bulge in the membrane to accomodate a fully inserted, rather than transbilayer, conformation. So, although the simulation does produce this transbilayer conformation, we have no reason to believe that this conformation would be distinct from the fully inserted conformation in biological systems. On the other hand, this configuration does exist in our model. Here, it is the reduction in the prevalence of this configuration that makes the other two configurations more prevalent for A$\beta 42$ than A$\beta 40$.

\subsection{FAD A$\beta$ peptide insertion}
\subsubsection{FAD mutations}

There are a number of known FAD mutations, including some involving A$\beta$ (as well as others involved in other aspects of the disease including A$\beta$ production). These are named by the populations they were first found in and include Flemish (A21G); Arctic (E22G); and Iowa (D23N)~\cite{kn:Murakami}. Murakami et al. also include Dutch (E22Q) and Italian (E22K) but there is some dispute about whether these are properly to be considered AD mutations~\cite{kn:Wattendorff, kn:Nilsberth, kn:Melchor}. To understand this, it is important to note that AD is often accompanied by cerebral amyloid angiopathy (CAA), deposition of A$\beta$ in blood vessels of the brain potentially leading to vessel rupture and stroke, especially in FAD cases~\cite{kn:Murakami}. ~\textcite{kn:Wattendorff} point out that classic Alzheimer's plaques are rarely found in the Dutch CAA case and dementia and death are due to cerebral hemorrhage involving damage to blood vessels, as is also the case in the Italian E22K mutant~\cite{kn:Melchor}. But the fact that AD also involves amyloid angiopathy leaves open the possibility that the Dutch and Italian forms are vascular forms of AD~\cite{kn:Wattendorff}.

Here, we set aside the issue of whether or not the Dutch and Italian A$\beta$ mutations are actually AD mutations or whether they should be regarded as something different and simply model insertion of these peptides into cell membranes. It is known that even the Dutch E22Q and Italian E22K mutant peptides interact with cell surfaces of cerebralvascular smooth muscle cells and cause cell death \emph{in vitro}~\cite{kn:Melchor}. Thus, it is conceivable that the mechanism may be similar to that described by Lin et al.\ \cite{kn:Lin} and the type of cells being damaged may simply be different -- either cerebralvascular smooth muscle cells~\cite{kn:Melchor} or brain pericytes~\cite{kn:Verbeek} rather than neurons. For simplicity, we will call all of these FAD mutations.

It is also important to note that these A$\beta$ mutations are autosomal dominant~\cite{kn:Nilsberth} and, in addition to lowering the age of onset for AD compared to sporadic cases, cause AD in all subjects with the mutations who live long enough. Many of these mutations also lead to increased A$\beta$ levels, but that cannot be the sole cause, as at least the Arctic mutation~\cite{kn:Lashuel} leads to decreased levels. The cause of disease also cannot be simply due to increased propensity to form fibrils or aggregates, because at least the Flemish mutation does not increase fibril formation~\cite{kn:Murakami}.

\subsubsection{FAD mutation results}

Results for these mutations are shown in Fig. 5, 6, and 7, and presented in reduced form in Table 1. We find that the most consistent difference in results relative to wild-type, for both A$\beta 40$ and A$\beta 42$, is that four of the five FAD mutants increase the number of steps the peptide is in a conformation like that of Fig. 2(c). The E22G mutant, however, results in a huge decrease of this conformation. In the other four cases, our results can easily be understood by considering the changes in polarity and hydrophobicity the mutations involve.

To understand our results, consider the changes in polarity and hydrophobicity the mutations involve. The polarity and hydrophobicity values used in this model have been tabulated before~\cite{kn:Maddox} and will not be repeated here, but it is worthwhile to discuss the mutations briefly. The A21G mutant involves a decrease in hydrophobicity, hence should cause residue 21 to more strongly prefer to be in the upper head region and thus result in an increase in the prevalence of the partially inserted conformation (Fig. 2(c)). This is what we find, at least for A$\beta 42$ (for A$\beta 40$, the percentage stays about the same). The E22Q mutant involves a fairly large increase in polarity and slight increase in hydrophobicity, thus the polarity should dominate and cause residue 22 to strongly prefer the upper head region, as we find. The E22K mutant is quite similar, but involves an even larger increase in polarity, and should also cause residue 22 to prefer the upper head region, as we find. And the D23N mutant involves a similar increase in polarity and slight increase in hydrophobicity, so again the polarity should dominate and increase this conformation, as we find. On the other hand, the E22G mutant replaces a polar, charged residue with a neutral, nonpolar residue and thus residue 22 is now mostly indifferent. This results in the observed huge decrease in the partially inserted conformation.

On one hand, the different behavior of the E22G mutant could be taken as evidence that there is no consistent change in insertion behavior, and thus suggest that this may not be the toxicity mechanism, or at least that oligomerization, not insertion, is important. That is certainly a possible meaning of these results. However, we would suggest that the reader compare the conformation of Fig. 2(c) with the channel structure of \textcite{kn:Durell}, specifically in Fig. 4 of that paper. Note that there are two helices predicted, from residues 15-24, running near the upper head region, and from 25-40 or 25-42, inserted. In Fig. 2(c), we have a nearly helical region from around residues 14-24 or 25 that is near the upper head region, and one from 25-40 which hangs down inserted. Thus it resembles the predicted structure more than either of the other conformations. In the structure prediction, the part of the chain prior to residue 15 folds down inside the oligomer and forms a beta-barrel, which we obviously do not capture here because we do not have oligomers necessary to stabilize such a structure, and because the model is not designed to capture beta structure.

With the similarity of Fig. 2(c) to the predicted channel structure in mind, we form the hypothesis that the partially inserted conformations (2(c), or 3(a) and (b)) are more likely to form channels, simply because of their resemblance to the structure of the monomers making up the channels. Below, in the artificial mutants section, we test this hypothesis against some experimental data for artificial A$\beta$ mutants, and present predictions for further tests.

One might argue that the observed changes in the prevalence of different conformations are relatively small and thus would be unlikely to result in a large difference in the toxicity of the different mutants. However, these channels consist of eight or twelve monomers. If one thinks of a large number of monomers inserted in a bilayer, some will be in each conformation at any given time, so we can think of the concentration of each conformation. Consider, then, the concentration of the conformation that can form channels. The time it takes for twelve monomers within a bilayer to find each other, or the probability, should scale as $c^{12}$, where $c$ is this concentration. Thus a small difference in the amount in the proper conformation could make a big difference in the likelihood of forming channels.

\subsection{Insertion of other mutant A$\beta$ peptides}

A variety of other data is available on mutant A$\beta$ peptides. Some reduce aggregation \emph{in vitro}~\cite{kn:Wurth}, some might cause FAD~\cite{kn:Janssen}, and some have various effects on cultured cerebrovascular smooth muscle cells~\cite{kn:Melchor, kn:VanNostrand}. Here, we examine some of the mutations which have been mentioned in the literature as a test of our hypothesis -- if it is right, and toxicity depends in part on the relative prevalence of the partially inserted conformation, our results ought to correlate with the experimental toxicity measurements.

First, we address the E22Q,D23N A$\beta 40$ double mutant created by \textcite{kn:VanNostrand}. This mutation's effect has been examined on human cerebrovascular smooth muscle cells (HCSM cells) because the E22Q mutation causes especially pronounced cerebral amyloid angiopathy and patients with this mutation typically die of hemorrhage, as discussed above. HCSM cells are known to degenerate in CAA in a manner that is associated with A$\beta$ deposition~\cite{kn:Melchor}. \textcite{kn:VanNostrand} found that the E22Q,D23N double mutant is even more toxic to HCSM cells than E22Q or D23N alone. Our idea was that the mechanism for this toxicity also involves insertion of the peptide and formation of channels, so we modeled this mutant as well. Results for this mutant are shown in Table 2; we find that the prevalence of the partially increased conformation does increase relative to wild-type. Relative to the D23N mutant, which we would predict would be the more toxic of the E22Q and D23N mutants, we observe an increase in the partially inserted conformation only for A$\beta 42$. Again, however, we would argue simply based on the residue properties that since both the E22Q and D23N mutants increase preference of those residues for the upper head region, the double mutant should have a stronger effect on this than either alone, and thus we expect that, given better statistics, we would agree with \textcite{kn:VanNostrand} and predict that the double mutant is more toxic than either alone.

\textcite{kn:Melchor} have found that an artificial E22D mutant of A$\beta 40$ does not effect HCSM cells, in contrast to biological E22Q and E22K mutants. They also observed that the A$\beta 40$ E22A mutant is toxic to HCSM cells. Therefore we model insertion of E22D and find (Table 2) that the E22D mutant results in a decrease in the partially inserted conformation (as one would expect due to the decrease in polarity), especially for A$\beta 40$. The E22A mutant results in a large decrease in the partially inserted conformation for both A$\beta 40$ and A$\beta 42$ (again, as one would expect due to the decrease in polarity). Thus, based on our hypothesis, we would agree that the E22D A$\beta 40$ mutant would not be toxic, but disagree that the E22A mutant would be toxic. This could be taken as evidence against our hypothesis or evidence that the toxicity mechanism is different for HCSM cells.

\textcite{kn:Janssen} recently identified a previously unknown mutation in the A$\beta$ peptide in two early-onset AD patients in the same family. This mutation, H6R, produced ages of onset around 55. We have here tried this mutation (results in Table 2) and find that it produces insertion behavior that is within error bars of wild-type. Given the position of the mutation, this is what we would expect, as it is within the range of amino acids (1-14 or more) that are firmly anchored in or near the upper head region, where there is some water content. Replacing histidine with arginine, which is even more polar, does not have a strong effect on this as both try to remain where there is water, i.e. the surface of the upper head region. Thus, if toxicity depends only on insertion conformation, we would suggest that this is indeed \emph{not} a FAD mutation. Thus we would suggest testing toxicity of this mutant in cell culture, particularly as already done for A$\beta 42$ by ~\textcite{kn:Lin}.

Some \emph{in vitro} work has been done to find artificial mutants that can reduce aggregation of A$\beta 42$~\cite{kn:Wurth}. Unfortunately, this has not yet been extended to include A$\beta 40$. However, we selected some of the point mutations which are known to reduce aggregation of A$\beta 42$ and modeled the insertion of these. We tried A2S, F19S, I32V, I32S, and V36E. Results are shown in Table 2. We find that F19S and I32S strongly increase the prevalence of the partially inserted conformation for both A$\beta 40$ and A$\beta 42$. For F19S, this is due to the substitution of polar serine for strongly hydrophobic phenylalanine, causing residue 29 to prefer the upper head region; for I32S, the reason is similar. In this case, the conformation is actually different from normal in that residue 32 also sticks in the upper head region. In contrast, the I32V mutant does not result in a large change relative to wild-type, consistent with the relatively small change between isoleucine and valine. Unsurprisingly, the A2S mutant makes no change to insertion behavior (residue 2 is firmly in the upper head region, so changing it to polar makes little difference). The V36E mutant, however, as one might expect, drastically increases the number of steps that the peptide is transbilayer, as the polar and charged glutamic acid strongly prefers to be in an environment with more water. Thus it decreases the prevalence of the partially inserted and fully inserted conformations. Thus we would predict that, if our hypothesis is correct, F19S and I32S should be the most toxic, I32V and A2S should be comparable to wild-type, and V36E \emph{might} be much less toxic than wild-type. We say \emph{might} because this probably depends on the thickness of the bilayer; if the bilayer is too thick it will probably behave just like wild-type since the E36 could not reach the lower head region. Thus a testable experimental prediction of our hypothesis is that toxicity of these mutants would be related to their insertion behavior as just described.

We can make a second prediction which is simply based on the observed insertion behavior of A$\beta 42$. Looking at the insertion behavior of the A$\beta 42$ versions of the Wurth et al. mutants, we find that I32V, A2S, and possibly V36E insert more like natural A$\beta 42$, while F19S, I32S, and possibly V36E insert differently. The reduction in aggregation splits the group differently -- I32S aggregates most, then A2S, V36E, and F19S are similar to one another and intermediate, and I32V aggregates least~\cite{kn:Wurth}. So if fibrillar aggregation primarily causes toxicity, experiments looking at toxicity should see the latter grouping, while if insertion behavior is of much more importance, toxicity experiments should see the former grouping.

\section{Discussion and conclusions}

We have here presented work applying a model of peptide insertion to A$\beta$, a peptide implicated in Alzheimer's disease. Specifically, we have examined the effect of FAD mutations on the peptide's insertion behavior, with the idea that any successful hypothetical toxicity mechanism should be able to explain why FAD mutations are toxic. Thus if FAD mutants do not affect peptide insertion into, or oligomerization within, membranes, the ion channel toxicity mechanism proposed previously~\cite{kn:Durell, kn:Arispe, kn:Lin} is probably not relevant biologically.

What we find is that the FAD mutations do affect peptide insertion. Four of these five mutations involve an increase in polarity or decrease in hydrophobicity and thus cause the peptide to prefer (relative to wild-type) a conformation where those residues are in the upper lipid head region (i.e. Fig. 2(c)). It is interesting to note that a channel structure suggested previously (\textcite{kn:Durell}, see Fig. 4) has these residues laying along the surface of the bilayer. Thus we find that four of the five FAD mutations increase the resemblance to this configuration.

Based on this similarity, we develop the hypothesis that causing the peptide to hang up in the upper leaflet (the partially inserted conformation) facilitates formation of harmful channels. We test this hypothesis on several artificial mutations examined \emph{in vitro} and find that it can explain the change in toxicity of two and is wrong on a third. As a further test of our hypothesis, we can offer some testable predictions. For example, if the hypothesis is right, we would suggest that the F19S mutant would increase toxicity of A$\beta$ relative to wild-type. Additionally,  there are several mutations known to reduce aggregation of A$\beta 42$ that we predict would promote the insertion behavior that would facilitate channel formation. Thus, it would be simple to distinguish between the channel formation toxicity mechanism and the aggregation toxicity mechanism by looking at the toxicity of these mutants. Even more evidence could be provided by replicating the work of \textcite{kn:Lin} but using various FAD mutants and looking at how these effect the abundance of channels. Additional information could also be gained from theoretical work along the lines of that by \textcite{kn:Durell} to see what effect these FAD mutants would have on channel structures.

Overall, this approach of modeling peptide insertion provides a simple way of making concrete predictions to distinguish the proposed mechanism of channel formation from others. The limitation of this approach, however, is that we can only look at insertion of single peptides, and not interaction between these. This is fine if formation of channels depends on having peptides initially in the correct conformations, which seems reasonable. However, it is also possible that interaction between inserted peptides causes them to adopt conformations appropriate for channel formation. If this is the case, then insight into channel formation would require a more sophisticated model that can include interaction between peptides. Even if this is not the case, including interaction between peptides will certainly answer the question of whether FAD mutants affect channel formation in a consistent way much more thoroughly than we are able to do here. Therefore, this would be a logical continuation of this work and is something we hope to do in the future.

Even if our hypothesis proves to be wrong, we have shown that the FAD mutants affect insertion behavior of the A$\beta$ peptide into lipid bilayers, and provided an understanding as to why the FAD mutants affect insertion in the way they do.

In conclusion, our work shows that most FAD mutations have a significant effect on the insertion of the A$\beta$ peptide in lipid bilayers in this theoretical model, and this effect can easily be understood by looking at the change in polarity and hydrophobicity accompanying the mutations. The effect of FAD mutations on insertion has not been studied previously, and may be significant. Additionally, we offer a hypothesis based on promoting channel formation by causing peptides to insert less fully that can help explain toxicity of A$\beta$ FAD mutants, as well as several artificial mutants studied \emph{in vitro}. While this hypothesis is unproven, it is based on the observation that these peptides do insert into cell membranes and form ion channels~\cite{kn:Lin}, and similarities to theoretically predicted channel structures~\cite{kn:Durell}. We provide testable predictions based on this hypothesis. It should be simple for experimentalists to disprove this hypothesis, if it is false, or to offer additional evidence for it, if they follow the experimental suggestions we offer above. Additionally, our work suggests the value of further modeling work to describe the full formation of these channels, rather than just single-peptide insertion.

\begin{acknowledgments}
We are grateful to R. Lal and G. Millhauser for useful discussions, and W. Pickett for a donation of computer time. We acknowledge support of the National Science Foundation (IGERT Grant DGE-9972741 and Grant No. PHY99-07949) (DLM), the U.S. Army (Congressionally Directed Medical Research Fund, Grant NP020132) (DLC), the MRSEC Program of the NSF under award number DMR-0213618 (MLL), and the NIRT Program of the NSF under award number 0210807. MLL acknowledges the generous gift of Joe and Essie Smith for endowing part of this work. 
\end{acknowledgments}

\appendix
\section{Data Analysis}
\small

The basic output of the simulation is the binned number of steps -- essentially the frequency with which each residue is found at each z-coordinate. This can be plotted across all residues and illustrates, as in Fig. 4, that there are multiple conformations that the peptide can switch between.

In terms of data analysis, it is more useful to plot z-distributions of specific residues which can be used to distinguish between the conformations described above. For example, residue 40 has a well-defined peak in the lower head region that can be used to distinguish the transbilayer conformation from the other conformations. A plot of residue 40 z-distributions for various mutations is shown in Fig. 5.

It is somewhat more difficult to distinguish exactly how many steps the peptide spends in the fully inserted and partially inserted conformations ((b) and (c) of Fig. 2). To understand this, it is important to recognize that each conformation results in a peak for a given residue. That is, from Fig. 2, one can easily see that the position of residue 28 has three significantly different locations depending on which conformation the peptide is in. As it turns out, it also has a fourth, which is not very different from the third, as can be seen in Fig. 3. Unfortunately, these peaks tend to overlap with each other quite significantly -- that is, residue 28 can sometimes have similar locations whether it is in the partially inserted conformation or the fully inserted conformation. This means that, to extract the number of steps in a given conformation from the data, it becomes necessary to fit some function to the peaks and then calculate the number of steps from that.

To do this, we select residues where the peaks appear to be particularly well separated. For one, we choose residue 30 (for A$\beta 40$), because at that residue, the peaks corresponding to the two partially inserted conformations actually overlap -- that is, residue 30 has the same average location no matter whether it is residues 22-23 alone that are stuck in the upper head region, or 22-23 and 26-28. This means that we can extract the weight of conformation (c) from Fig. 2 -- the conformation grouping these two together -- by looking at this residue. For A$\beta 42$, the two peaks of the partially inserted conformation do not quite overlap for residue 30, but do for residue 35, so we do the same analysis, but for residue 35. A plot of the number of steps at each z-value for these residues is shown in Fig. 6. 

To separate the two partially inserted conformations of Fig. 3, we choose three residues that have four peaks that are the most well-separated (unlike residue 30 or 35, where these two peaks merge into one). These are residues 26, 28, and 31. Plots of the number of steps at each z-value for different mutations for one of these residues, residue 26, are shown in Fig. 7.

Our goal was to find the number of steps in each conformation. Having chosen the residues where the peaks are the best separated, it is necessary to find the number of steps under each peak. Since they overlap, we find that the best way to do this is to use the least-squares method to find fits to each peak that best describe the whole function. This is probably best understood using a concrete example, for which we choose residue 26.

For residue 26, we know there are four peaks, and, as seen in Fig. 7, the locations of these are fairly clearly visible. To fit these peaks, we begin by averaging all of our results for the wild-type form of A$\beta 40$. We assume that the peaks are Gaussian. Since the peaks are reasonably well separated, we assume initially that the center of each Gaussian is at the maximum. Then, we perform a least-squares fit of the standard deviation and amplitude of each Gaussian. Particularly, we perform our fit by looping through the peaks and suggesting changes first to the standard deviation and then to the amplitude. Each time, we try both increasing and decreasing the standard deviation by a specified step size, and check whether this improves the quality of the fit. We then move on to the next peak and do the same thing, then repeat the process for the amplitude. Then we reduce the step size and repeat the whole process. We do this until the fit can no longer be improved by further iterations of the process. Having done all that, we also then try altering the locations of the centers of the Gaussians slightly to see if it improves the fit. Having then selected optimal standard deviations and center locations, we store those, and assume that the shape and location of each peak will remain the same for other simulations, simply the amplitudes will vary. A sample such fit is shown in Fig. 8. As is apparent from the figure, the fit is good. Therefore, we concluded that assuming the peaks are Gaussian is sufficient, at least to get a reasonable estimate of the number of steps in each conformation.

\begin{figure}
\includegraphics{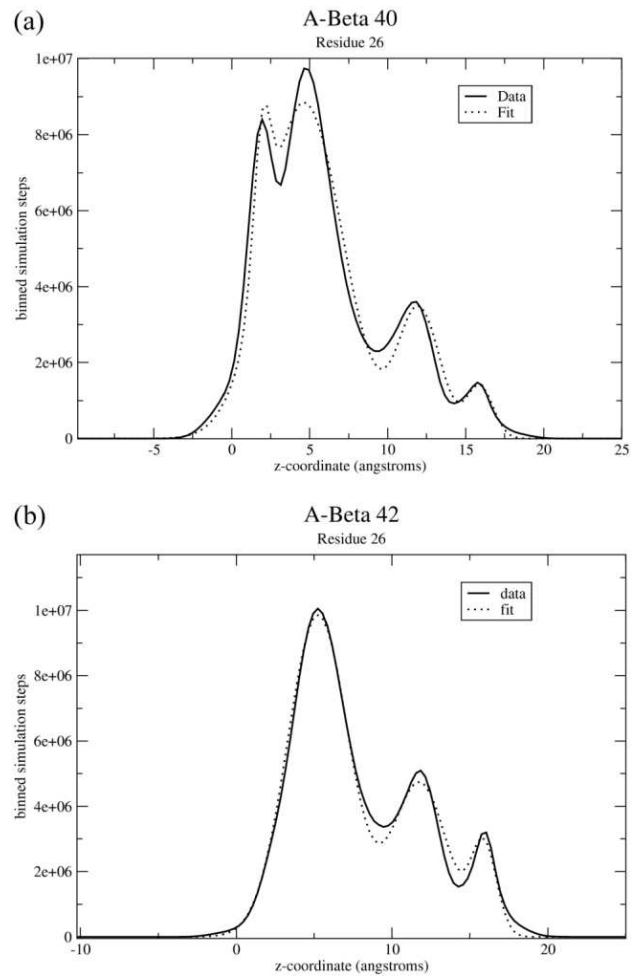}
\caption{{\bf Binned number of steps spent at each z-coordinate for residue 26 of (a) A$\beta 40$ and (b) A$\beta 42$, along with fits used for data analysis}. Shown here are the data for wild-type, compared to fits of Gaussians, as described in the data analysis section. The Gaussians appear to provide fairly good fits and thus we use these in this work to calculate the number of steps in each conformation.}
\end{figure}

Having done this, we are able to then perform a four parameter fit for every other data set for every A$\beta 40$ mutation data set, simply by fitting the amplitudes while keeping the standard deviations and peak locations fixed. We have also tried allowing the standard deviations to vary when doing these subsequent fits but this does not substantially improve the quality of the fit, and it seems reasonable that the shape of a given peak for a given residue should be constant.

We apply a similar technique to fitting residues 28 and 31, and residue 30, but we fit residue 30 using only three peaks, and we do the same for A$\beta 42$. Additionally, using this technique, we are able to calculate standard deviations for a given peptide or mutant. For example, for the D23N mutant, where residue 23 is changed from aspartate (D) to asparagine (N), we perform a fit separately for each of our five MC runs and calculate the area under each peak. Then, we average the results over all five trials and calculate the standard deviation.

For the four peak case, it is somewhat difficult to accurately separate the third and fourth peaks, thus it helps that we are able to separately perform fits to four peaks on residues 26, 28, and 31. We are thus able to take the apparent change of steps under each of these three peaks relative to the wild-type and average over all three residues. This significantly reduces our error.

There is one more factor which complicates issues. Going back to our earlier example of the location of residue 28 in Fig. 2, it appears in very different positions along the z-axis depending on which conformation the peptide is in. However, it is also possible to have conformations where residue 28 may, for example, be quite low, like it is in the transbilayer conformation (Fig. 2(a)) but residues 38-40 are not in the lower leaflet. This effect gets worse the further the residue being examined is from residue 40, and is apparent in Fig. 6 and 7 as the area under the ``transbilayer'' peak on those residues is obviously much larger than the area under the same peak on residue 40, simply because more conformations look similar to the transbilayer conformation.

This means that the calculated number of steps under each peak for residues 26, 28, 30, and 31, while accurate, are not really to be taken as a measure of how many simulation steps the peptide spends in each of those conformations. Since the peak shapes remain constant over the different mutations, however, these \emph{do} accurately describe the change relative to the wild-type. That is, an increase in the prevalence of the conformation of Fig. 2(c) relative to wild-type for a given mutation is correct, while the absolute number of steps in that conformation may not be accurate, except for the conformation of Fig. 2(a), which we can extract accurately from residue 40.

As we discussed above, for many of the peptides we have studied we have averaged over five trials. However, for some we have used significantly more trials. Specifically, for wild-type of A$\beta 40$ and A$\beta 42$, we used more trials, as it was particularly important to have a good average for those results since we compare all of our other results to those. For A$\beta 40$ we used 25 trials; A$\beta 42$ we used 20. Additionally, for mutants with particularly small changes relative to wild-type, or particularly large standard deviations, we also used more runs. We did 20 trials for A21G A$\beta 40$; 10 for A21G A$\beta 42$, and 10 each for A$\beta 42$ E22Q and D23N. Of the artificial mutants, we did 10 trials for A$\beta 40$ and A$\beta 42$ of both I32S and H6R.



\end{document}